\documentclass[aps,prapp,twocolumn,superscriptaddress,longbibliography]{revtex4-1}

\usepackage{graphicx}
\usepackage{mathtools}
\usepackage{verbatim}

\begin{document}

\title{Tunable Super-Mode Dielectric Resonators for Axion Dark Matter Haloscopes}

\author{Ben T. McAllister}
\email{ben.mcallister@uwa.edu.au}
\address{ARC Centre of Excellence for Engineered Quantum Systems, School of Physics, The University of Western Australia, Crawley 6009, Australia}
\author{Graeme Flower}
\address{ARC Centre of Excellence for Engineered Quantum Systems, School of Physics, The University of Western Australia, Crawley 6009, Australia}
\author{Lucas E. Tobar}
\address{ARC Centre of Excellence for Engineered Quantum Systems, School of Physics, The University of Western Australia, Crawley 6009, Australia}
\address{Department of Electrical and Computer Systems Engineering, Monash University, Clayton 3800, Australia}
\author{Michael E. Tobar}
\email{michael.tobar@uwa.edu.au}
\address{ARC Centre of Excellence for Engineered Quantum Systems, School of Physics, The University of Western Australia, Crawley 6009, Australia}

\date{\today}

\begin{abstract}
We present frequency tuning mechanisms for dielectric resonators, which undergo ``super-mode" interactions as they tune. The tunable schemes are based on dielectric materials strategically placed inside traditional cylindrical resonant cavities, necessarily operating in Transverse Magnetic modes for use in axion haloscopes. The first technique is based on multiple dielectric disks with radii smaller than that of the cavity. The second scheme relies on hollow dielectric cylinders similar to a Bragg resonator, but of different location and dimension. In particular we engineer a significant increase in form factor for the TM$_{030}$ mode utilising a variation of a Distributed Bragg Reflector Resonator. Additionally, we demonstrate application of traditional Distributed Bragg Reflectors in TM modes, which may be  applied to a haloscope. This is the first demonstration of Bragg resonators applied to TM modes, as well as the first application of super-modes to tune Bragg resonators, or haloscope resonators. Theory and experimental results are presented showing an increase in Q-factor and tunability due to the super-mode effect. The  TM$_{030}$ ring resonator mode offers between 1 and 2-orders-of-magnitude improvement in axion sensitivity over current conventional cavity systems and will be employed in the forthcoming ORGAN experiment.
\end{abstract}


\maketitle
\section{Introduction}
For many years scientists have suspected the presence of a large amount of so called ``dark matter" in the galactic halo, motivated by a number of gravitational observations. Despite this, the nature of this matter remains unknown, although many have suggested that it may be composed of new particles not included in the standard model. One popular dark matter candidate is a weakly-interacting slim particle (WISP)~\cite{wisps} known as the axion. Axions were first proposed in 1977 as a consequence of an elegant solution to the strong CP problem in QCD~\cite{PQ1977}. Since the expected properties of the axion (finite mass with weak coupling to regular matter) align with the desired properties of dark matter, it was proposed in 1983 that dark matter might be composed of axions~\cite{Sikivie1983b}.

The most mature and common laboratory search technique for axions is known as the haloscope, which was first proposed by Sikivie in 1983~\cite{Sikivie83haloscope,Sikivie1985}. The haloscope aims to detect axions via their coupling to photons. It is thought that axions will convert into photons in the presence of other photons (a form of the Primakoff effect). In most haloscopes a strong, external static magnetic field provides a source of virtual photons for axions to scatter off and create real photons. Due to conservation of energy the frequency of the generated real photon corresponds directly to the mass of the axion (with some narrow linewidth as a result of velocity dispersion). Many axion haloscopes are currently operational, most notably the Axion Dark Matter Experiment (ADMX), the first and most mature such experiment~\cite{ADMXaxions2010,ADMX2011}.

A confounding concern for axion haloscopes is the fact that the mass of the axion is largely unknown (other than some broad cosmological limits~\cite{Sikivie1983,Preskill1983}), meaning that the frequency of the generated photons is also unknown. Additionally the strength of the axion coupling to photons is unknown, this creates a large parameter space for searching. The critical parameter, which haloscopes ultimately wish to constrain or bound is the Peccei-Quinn symmetry breaking scale, $f_a$. This determines both the axion mass and the strength of its coupling to photons according to\\
\begin{align}
	\text{m}_a&\sim\frac{4.51\times10^{15}}{f_a}~\text{eV}\nonumber \\
	\text{g}_{a\gamma\gamma}&=\frac{\text{g}_{\gamma}\alpha}{f_{a}\pi}.\nonumber
\end{align}
Here $\text{g}_\gamma$ is an axion-model dependent parameter of order 1, and $\alpha$ is the fine structure constant~\cite{K79,Kim2010,DFS81,SVZ80,Dine1983}.

\section{Haloscopes and dielectric materials}
In a haloscope, a resonant cavity is embedded in a strong static magnetic field. If axions are present due to an abundance in galactic halo dark matter, a small number will convert into real photons with a frequency corresponding to the axion mass. It is advantageous to tune the resonant frequency of the cavity to the corresponding photon frequency in order for the signal to be resonantly enhanced. Once the signal is trapped in the cavity, it can be read out via low noise electronics. The expected signal power in a haloscope is given by~\cite{Daw:1998jm}\\
\begin{equation}
\text{P}_a\propto\text{g}_{a\gamma\gamma}^2B^2CVQ_L\frac{\rho_a}{\text{m}_a}\nonumber
\label{eq:Paxion}.
\end{equation}
Where $B$ is the field strength of the external magnetic field, $V$ is the volume of the detecting cavity, $Q_L$ is the loaded cavity quality factor (provided it is lower than the expected axion signal quality factor $\sim10^6$), $\rho_a$ is the local axion dark matter density, and $C$ is a mode dependent form factor of order 1~\cite{McAllisterFormFactor}, defined generally in dielectric and magnetic materials as
\begin{equation}
	\text{C}=\frac{\left|\int dV_{c}\vec{E_c}\cdot\vec{\hat z}\right|^2}{2~V\int dV_{c}\epsilon_r\mid E_c\mid^2}+\frac{\frac{\omega_a^2}{c^2}\left|\int dV_{c}\frac{r}{2}\vec{B_c}\cdot\vec{\hat\phi}\right|^2}{2~V\int dV_{c}\frac{1}{\mu_r}\mid B_c\mid^2}.\nonumber
\end{equation}
Here $E_c$ and $B_c$ are the cavity electric and magnetic fields respectively, and $\epsilon_r$ and $\mu_r$ are the relative dielectric and magnetic constants of the media. It is worth noting that the two terms in this equation are equal to one another, and we may present it as
\begin{equation}
\text{C}=\frac{\left|\int dV_{c}\vec{E_c}\cdot\vec{\hat z}\right|^2}{V\int dV_{c}\epsilon_r\mid E_c\mid^2}.
\label{eq:C}
\end{equation}
\begin{figure*}\centering
	\includegraphics[width=\textwidth]{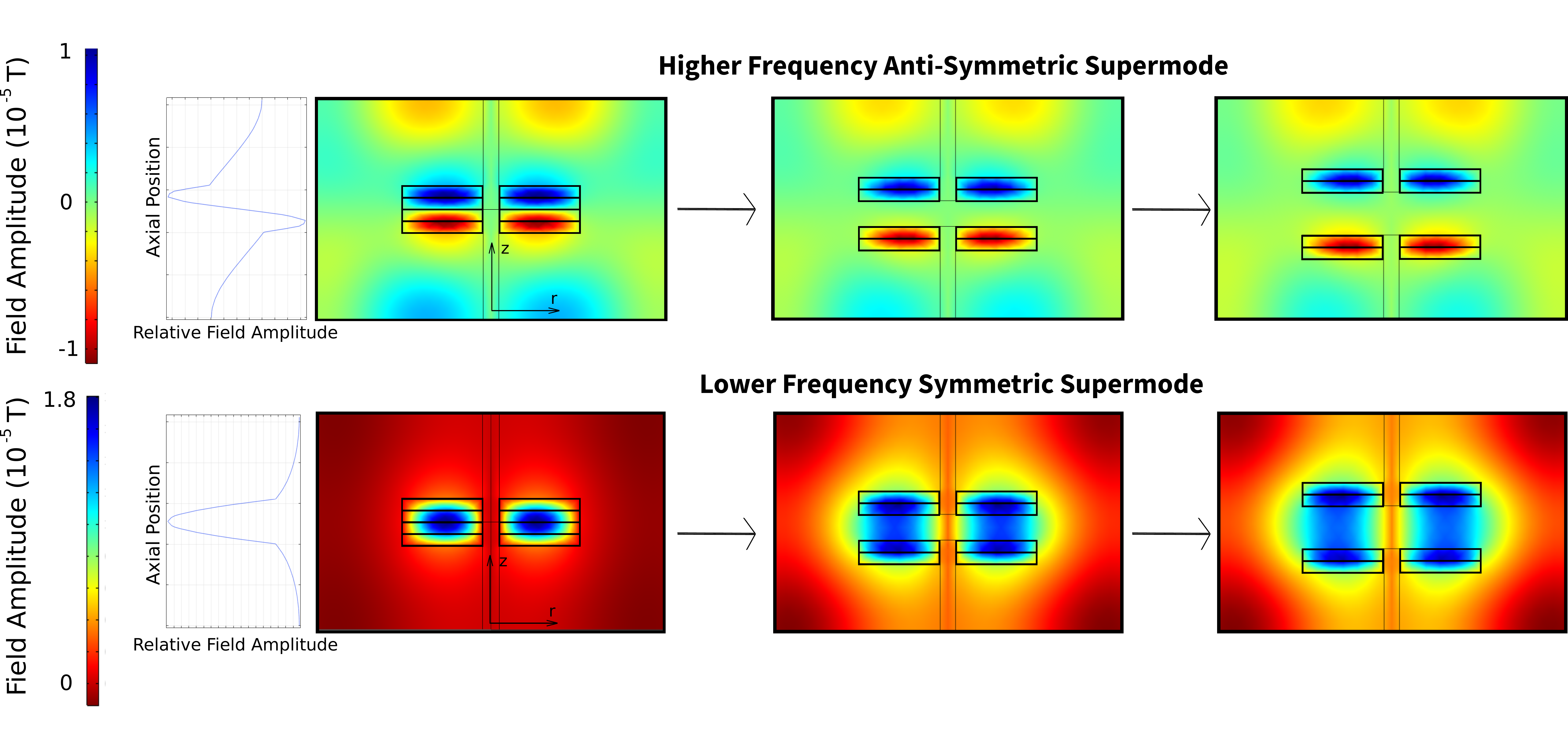}
	\caption{$B_\phi$ field distribution for an antisymmetric and symmetric super-mode pair. The plots on the left show B$_\phi$ as a function of z for fixed radius, whilst the axes show negative and positive field values in Tesla. Both representations are computed in COMSOL.  As the gap is increased the lower mode (TM$_{012}$-like) can be identified as having in-phase lobes in the two dielectric pieces (symmetric super-mode), while the upper mode (TM$_{013}$-like) has similar lobes, which are out of phase (anti-symmetric). In the limit that the gap spacing is large the frequency of the symmetric mode approaches the frequency of the anti-symmetric mode, depending on the mode confinement within the dielectric. The symmetric super-mode is sensitive to axions as the field is largely in phase across the volume, while the anti-symmetric mode is not. The sapphire disks are represented by heavy dark lines.}
	\label{fig:BmodesDisk}
\end{figure*}
We could equally present it as twice the second term in the full equation, dependent on B$_\phi$. The scanning rate of a haloscope is given by~\cite{Stern:2015kzo}\\
\begin{equation}
\frac{df}{dt}\propto\frac{1}{\textit{SNR}_{goal}^2}\frac{\text{g}_{a\gamma\gamma}^4B^4C^2V^2\rho_a^2Q_LQ_a}{m_a^2(k_BT_n)^2}.
\end{equation}
Here $\textit{SNR}_{goal}$ is the desired signal-to-noise ratio of the search, $Q_a$ is the axion signal quality factor, and $T_n$ is the effective noise temperature of the first stage amplifier, with later amplifier contributions suppressed by the gain of this amplifier. This is the quantity that must be maximized in design of an experiment, for which $C^2V^2G$ can be viewed as a figure of merit for resonator design, as these are the only resonator dependent terms. G is the mode geometry factor given by\\
\begin{equation}
G=\frac{\omega\mu_0\int{|\vec{H}|^2dV}}{\int{|\vec{H}|^2dS}},
\nonumber\end{equation}
which is directly proportional to the mode quality factor according to
\begin{equation}
Q=\frac{G}{R_s},
\nonumber\end{equation}

where $\omega$ is the angular resonant frequency, $\mu_0$ is the permeability of free space, $\vec{H}$ is the cavity magnetic field, and $R_s$ is surface resistance of the material. In this calculation there is the implicit assumption that any dielectric adds no loss. This is valid when using low-loss materials such as sapphire and rutile etc.~\cite{krupka99,krupka1999b,tobar98}, as the loss produced at the cavity walls will usually be orders of magnitude greater for Transverse Magnetic (TM) modes with zero azimuthal mode number. Otherwise, one must compute the filling factor in the dielectric and multiply by its loss tangent to calculate the effect on the Q-factor.

Recently, there has been much interest in the use of dielectric materials in axion haloscopes~\cite{Orpheus,MADMAX,ADMXHF2014}. There are many reasons for this interest. Firstly, as evident in the form factor expression given by eq.~\ref{eq:C}, the relative dielectric constant of a medium has an impact on the coupling of the axion to the cavity resonance, and this can be exploited to boost the form factor with careful placement of dielectrics. Secondly, dielectric resonators are well known for their high quality factors, and are often employed in other areas, such as metrology, where high quality resonances are required. Additionally, the introduction of dielectrics into a cavity provides many more free parameters and opportunities for broken symmetry that can be exploited to create a frequency tuning mechanism.

Furthermore, as there is increased interest in axion searches in mass ranges above and below the range traditionally searched by cavity experiments~\cite{McAllister:2016fux,Sikivie2014a,ABRACADABRA,CULTASK,YaleAxion,ORGANPatras2016,SMASH}, dielectrics have been of increased interest, due to their ability to lower the resonant frequency of a cavity by lowering the speed of propagation of resonant photons, or increasing the effective optical path length.

In the push towards higher frequency and mass searches, it would be beneficial to utilize high order resonant modes, as they can provide high frequency resonances in large resonator volumes. However, in a traditional empty resonator higher order resonances have significantly reduced form factors, due to a high degree of field variation. For example, comparing a TM$_{010}$ mode with a TM$_{020}$ mode in the empty cavity, the lower order mode has all of the E$_z$ field in the same direction, and thus the form factor is high ($\sim$0.69). In contrast, the higher order mode has an out of phase E$_z$ field component, which `cancels out' part of the coupling and reduces the form factor to $\sim$0.13. As we will discuss, careful placement of dielectric materials can assist with this problem. Moreover, the modes in the empty cavity resonator have frequencies which are independent of length, and thus elaborate tuning mechanisms must be implemented~\cite{Stern:2015kzo}. We show in the proceeding section that the concept of super-mode tuning allows altering resonator frequency by simply varying positions of two components of the resonator along the length of the cylinder.

\section{Dielectric Resonator Proposals}
\begin{figure*}[t]
\centering
	\includegraphics[width=0.9\columnwidth]{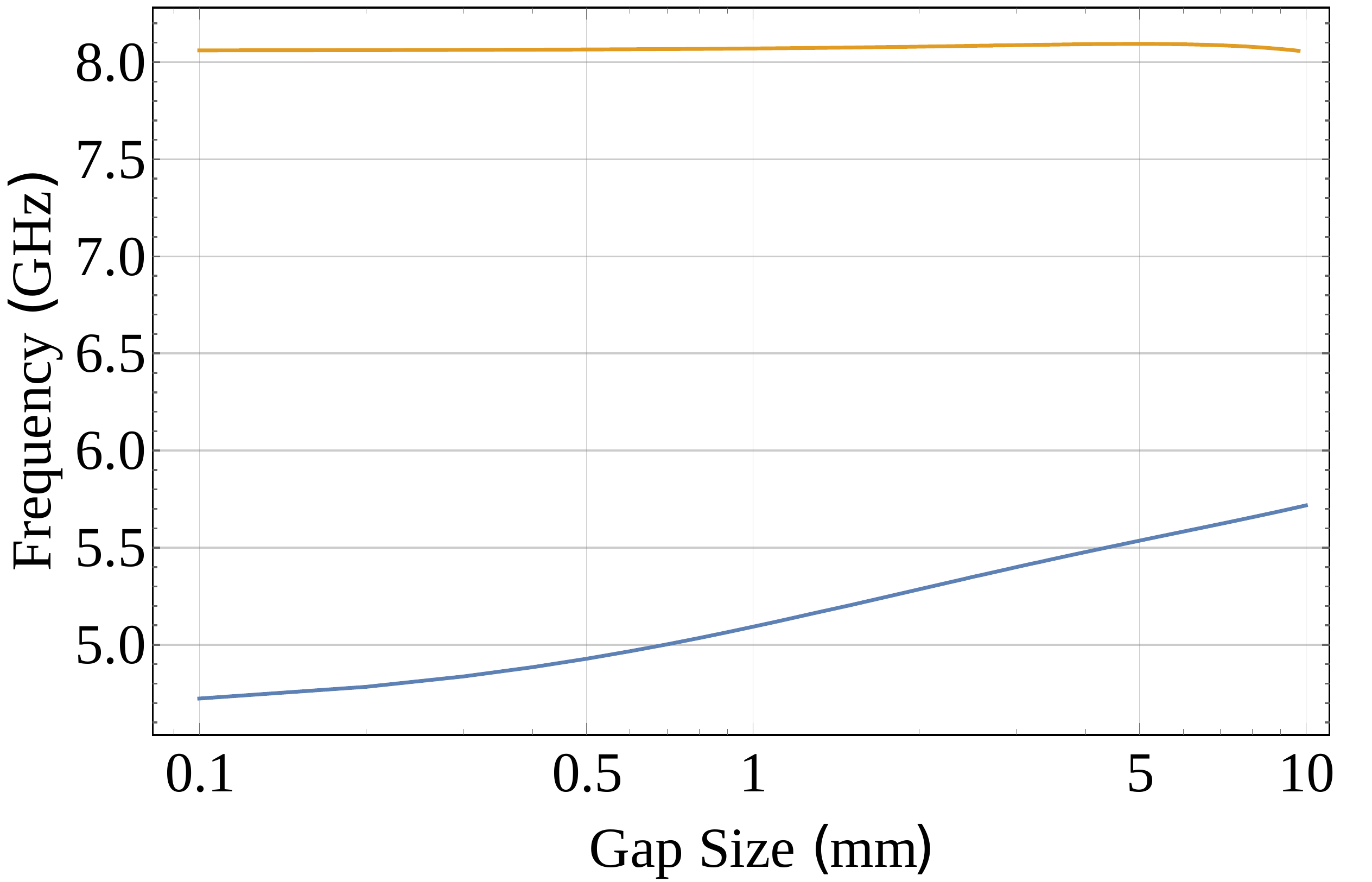}~
	\includegraphics[width=0.9\columnwidth]{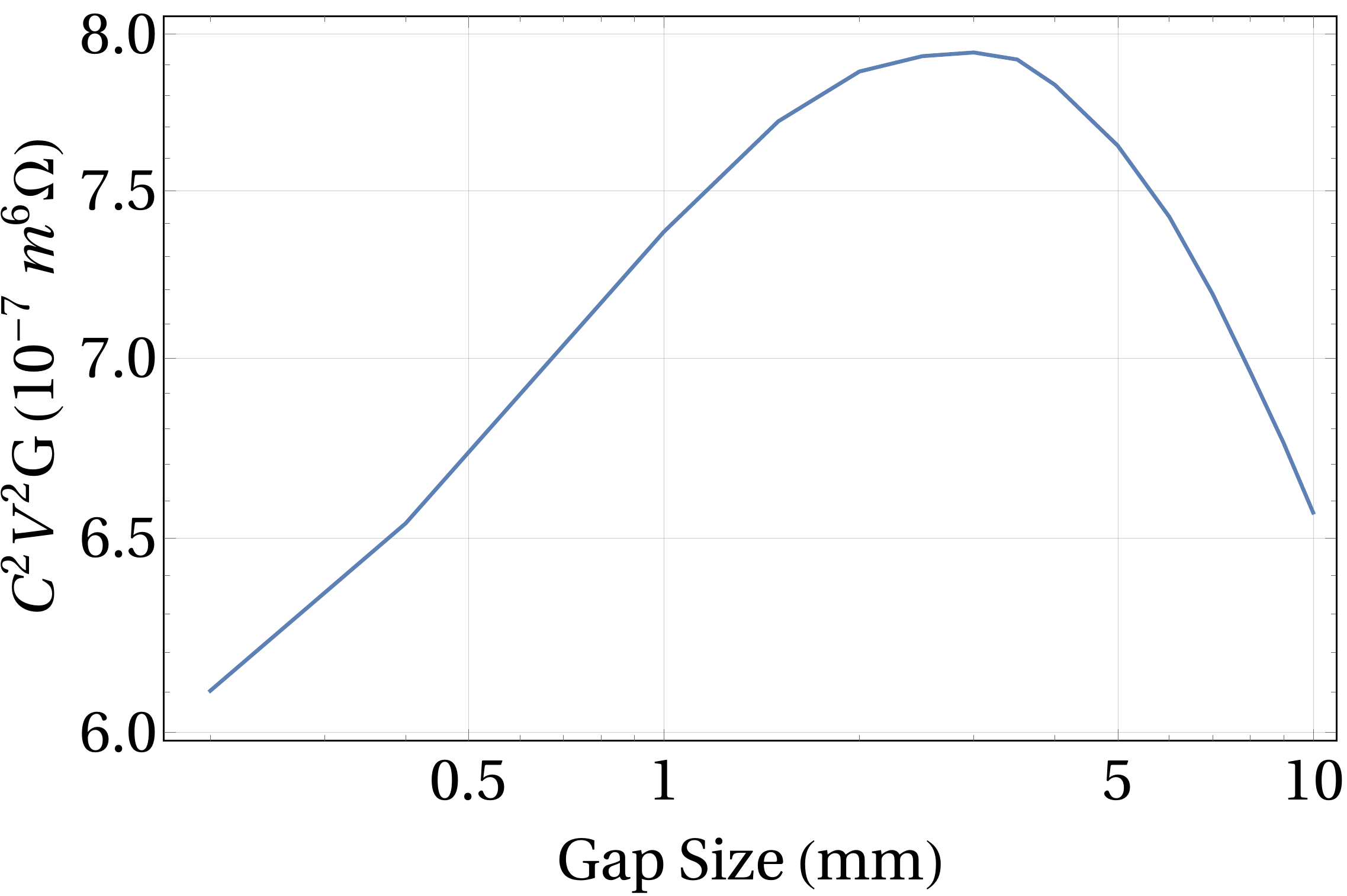}
	\caption{Calculated frequencies in GHz for the symmetric and anti-symmetric super-modes (left, blue and orange respectively), and $C^2V^2G$ product for the symmetric super-mode (right) vs gap size in millimetres for the disk resonator discussed in the text.}
	\label{fig:SupermodeCVG}
\end{figure*}
\begin{figure}\centering
	\includegraphics[width=0.9\columnwidth]{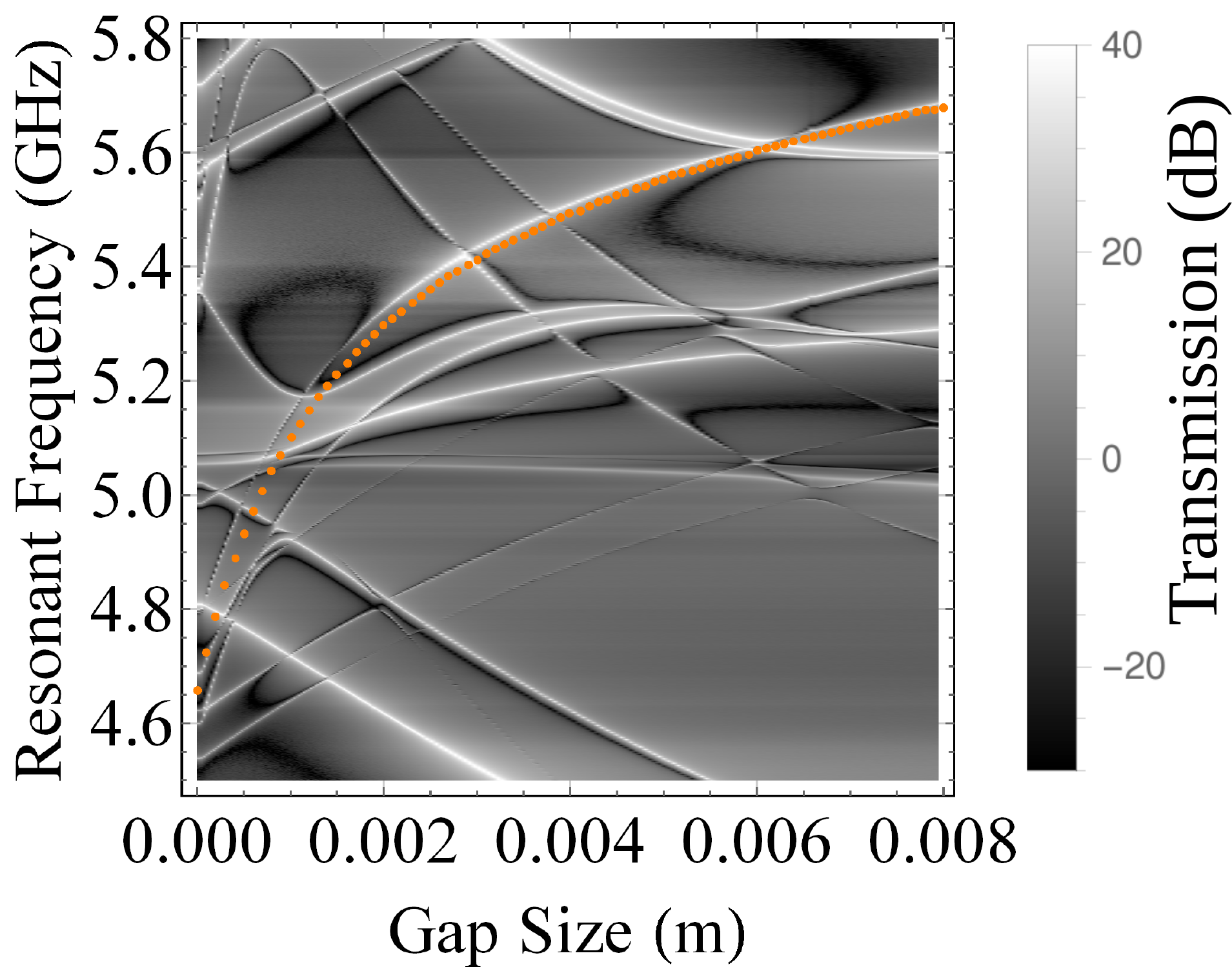}
	\caption{Density plot of transmission measurements in the proof of concept experiment. Lighter colours represent more transmission, whereas darker colours represent less transmission. The orange overlaid data represents the frequency of the most sensitive mode vs gap size computed via finite element analysis. We attribute the frequency discrepancy at small gap sizes to a misalignment between the two disks, which has less impact on the modes as the gap size increases. We observe multiple mode-crossing with disimilar modes, whilst the higher frequency anti-symmetric super-mode is too high in frequency (roughly 8 GHz as per fig.~\ref{fig:SupermodeCVG}) to be observed on this plot.}
	\label{fig:supermodePoC}
\end{figure}
\begin{figure*}[t]
\centering
	\includegraphics[width=0.9\textwidth]{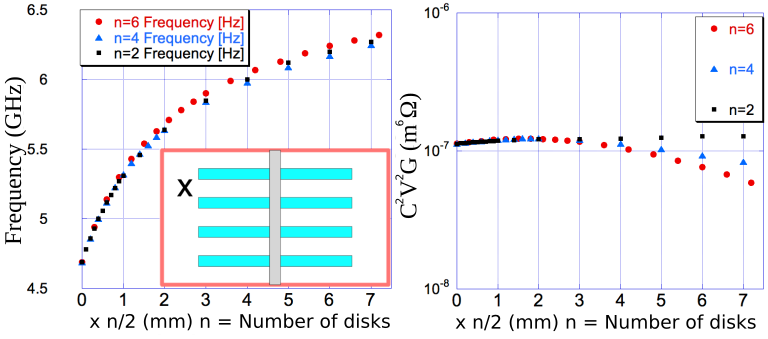}
	\caption{Comparison of super-mode frequency tuning (left) and sensitivity (right) as a function of $x\times n/2$. The inset on the left figure shows a schematic of a multi-disk resonator of $n=4$, with each disk seperated by $x$ mm.}
	\label{fig:multidisk}
\end{figure*}
The Electric Tiger and MADMAX experiments propose to utilize multiple dielectric disks or slabs and simultaneously tune multiple gap spacings~\cite{Orpheus,ET1,ET2}, with the former being a resonant system and the latter broad-band. Whilst sensitive and valuable proposals, they present some practical challenges. One potential problem with such modes in a resonant system is that they are of extremely high order and without proper design can suffer from mode crowding and require the discs to be larger than the spot size to maintain high-Q factor. In this work we propose a different technique, which utilises a super-mode tuning mechanism in a dielectric loaded cavity resonator. For this technique only two pieces of dielectric are required (utilizing more pieces is possible, but does not give any advantage) and like MADMAX and Electric Tiger the tuning is ``built-in", which is to say that we do not need to introduce extra material into the cavity to tune the modes.  We propose two different types of dielectric resonators which utilise this technique, and determine the sensitivity of each scheme to axions. We compare these results with a haloscope tuned by traditional means.

\subsection{Dielectric Disks}

The first scheme proposed here relies on multiple dielectric disks (with a minimum of two), located in a cylindrical conducting shell with a variable gap between them. The simplest dielectric modes that tune in frequency are known as super-modes \cite{supmode1,Vahala2010,cuth96}. In the two disk structure symmetric and anti-symmetric super-modes span both cylindrical disks of dielectric (in this case, sapphire) and occupy the entire volume. The magnetic field $B_{\phi}$ density plots are shown in fig.~\ref{fig:BmodesDisk}, computed using the finite element method in COMSOL Multiphysics. The higher frequency mode is the anti-symmetric mode with a zero in field between the two disks and each anti-node out of phase, while the symmetric mode is lower in frequency with two anti-nodes in phase. In the limit that the gap spacing is large the frequency of the symmetric mode approaches the frequency of the anti-symmetric mode, depending on the mode confinement within the dielectric, this effect can be seen in fig.~\ref{fig:SupermodeCVG}. The initial frequency difference between the symmetric and anti-symmetric modes defines the potential maximum possible tuning range using this technique. We note that these symmetric and anti-symmetric super-modes arise as perturbed version of the cavity TM$_{012}$ and TM$_{013}$ modes respectively. The analytical solutions for the B$_\phi$ components of TM$_{01p}$ modes are
\begin{equation}
\text{B}_\phi(r,z) \propto J'_0(\frac{\zeta_{0,1}}{R}r)\text{cos}(\frac{p\pi}{L}z)
\end{equation}
Where $\textit{J'}_0$ is the derivative of the 0th order Bessel J function, $\zeta_{0,1}$ is the 1st root of the 0th order Bessel J function, $R$ is the cavity radius, and $L$ is the cavity length. If we observe the B$_\phi$ mode profiles as a function of z for fixed r presented in fig.~\ref{fig:BmodesDisk} we can see that the symmetric mode conform to this field structure for a p value of 2, with some compression and distortion due to the presence of the dielectric, whilst the anti-symmetric mode conforms for a p value of 3, with similar compression and distortion.
This super-mode interaction can provide large frequency tuning ranges over small displacements. In the ideal situation the uppermost disk would be displaced upwards whilst the lower disk is displaced downwards to maintain symmetry. In practice it is more feasible to leave one disk stationary and adjust the position of the other relative to it. Over a short tuning range, this is not a large deviation from the ideal. For the specific parameters which were modelled for our study, in order to achieve tuning of roughly 500 MHz from the starting frequency of 4.66 GHz, it was only necessary to increase the gap between the disks by a small fraction (2\%) of the length of the resonator (in this case 1 mm). The $C^2V^2G$ product and resonant frequencies as a function of gap size for this example (cavity radius $\sim$41 mm, height $\sim$50 mm, sapphire radius 21 mm, sapphire height 2.74 mm, central teflon post radius of 2mm,  and a symmetric super-mode starting frequency of 4.66 GHz) are shown in fig.~\ref{fig:SupermodeCVG}. These dimensions were chosen due to availability of materials, but the resonator may be arbitrarily scaled to provide any desired frequency.

We performed proof of concept measurements to verify the modes behaved as expected. A resonator with the above dimensions was constructed, and the disks were placed on teflon rods. The lower disk remained stationary whilst the upper disk was displaced upwards using a micrometer, this of course yields slightly different results than those presented in fig.~\ref{fig:SupermodeCVG}. Two probes were inserted into the cavity to read out the relevant modes in transmission. Expected and measured frequencies as a function of gap size are presented in fig.~\ref{fig:supermodePoC}.

This resonant structure is novel as it contains a ``built-in" highly responsive frequency tuning mechanism and is readily scalable to different frequencies. Furthermore, it is a relatively simple structure compared with other proposed resonators that rely on many more dielectric disks or slabs, and will be more practical to implement experimentally. Particularly the responsiveness of the resonant frequency to position displacement is appealing, combined with the fact that spurious mode density in the region is very low. 

We decided to investigate the effect of tuning multiple disks (where $n =$ the number of disks) in our cavity resonator by implementing a finite element model in COMSOL. Results are shown in fig. \ref{fig:multidisk}. In this case we implement a copper cavity of radius 31mm and height 37mm with multiple disks supported by a central sapphire rod of 3mm in diameter. Here, we assume that the amount of sapphire remains constant in each simulation, i.e. for $n=2 $ the individual disks height is 6mm, for n=4 the individual disks height is  3mm and for n=6 the individual disks height is 2 mm. The equivalent super-mode behaves in a very similar way in all cases, as can be seen in fig. \ref{fig:multidisk}, where we show frequency and $C^2V^2G$ versus $x\times n/2$, where $x$ is the individual gap between two adjacent disks. Thus, for simplicity (and for sensitivity) the $n=2$ choice is optimum for this type of tuning. Introducing more disks will lead to greater technical difficulties in simultaneously tuning all disks, and does not yield an increased sensitivity. Higher order modes equivalent to those used by Electric Tiger and MADMAX will exist, however in a closed cavity the density of modes would be too great, these type of modes are best suited to open resonators to avoid mode crowding. To keep the high Q-factor of the super-modes, the cavity must be present as the relevant modes are of low azimuthal order, and thus modes would radiate a significant amount of energy without a cavity and exhibit a degraded Q-factor.

\subsection{Dielectric Rings}

\begin{figure*}\centering
	\includegraphics[width=0.75\textwidth]{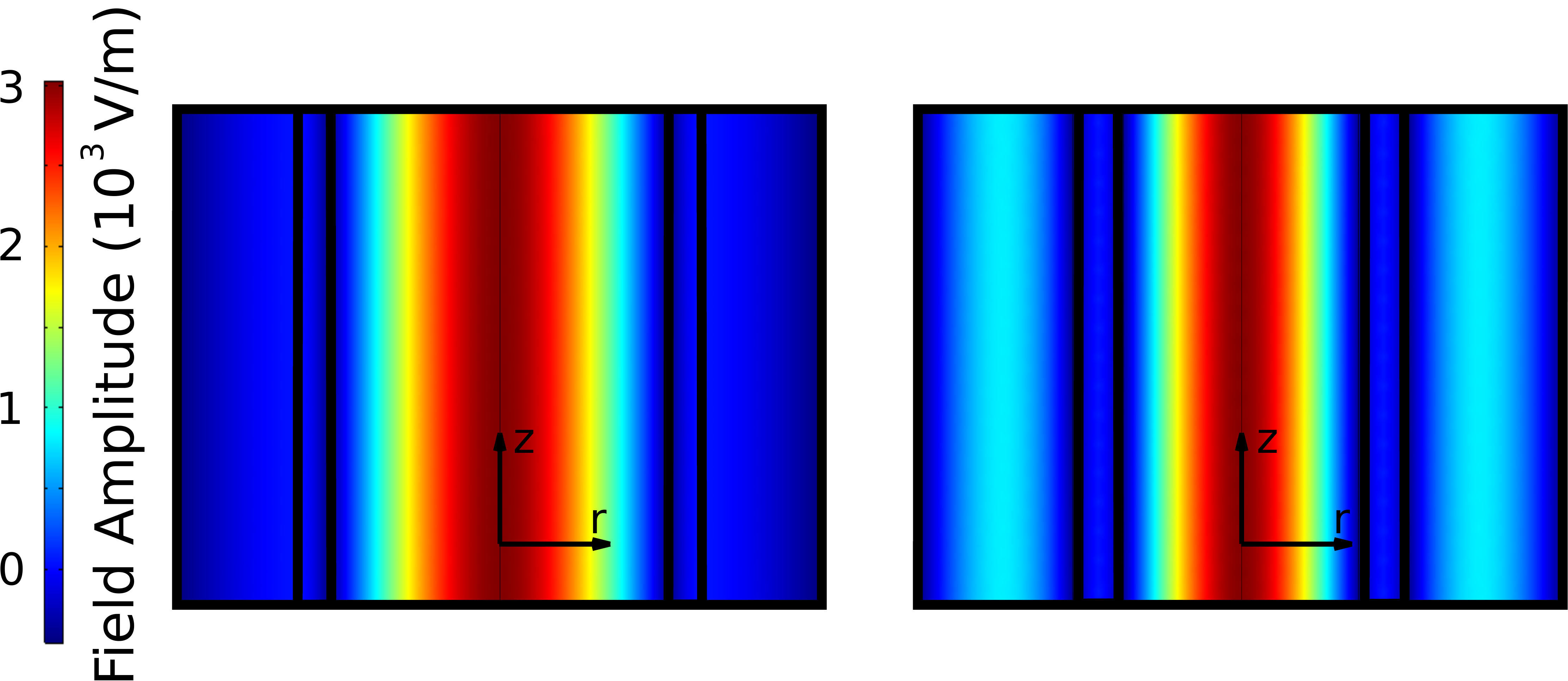}
	\caption{A 2D axisymmetric visualization of the z-component of electric field for a TM$_{020}$-like mode confined inside a sapphire ring, such that the Bragg effect is achieved (left), and a similar visualization of a TM$_{030}$-like mode designed such that the out of phase electric field is contained in the sapphire ring and suppressed. We name this phenomenon the Dielectric Boosted Axion Sensitivity (DBAS) effect (right). The sapphire rings are represented by heavy dark lines.}
	\label{fig:ClassicalBragg}
\end{figure*}

\begin{figure*}\centering
	\includegraphics[width=0.9\columnwidth]{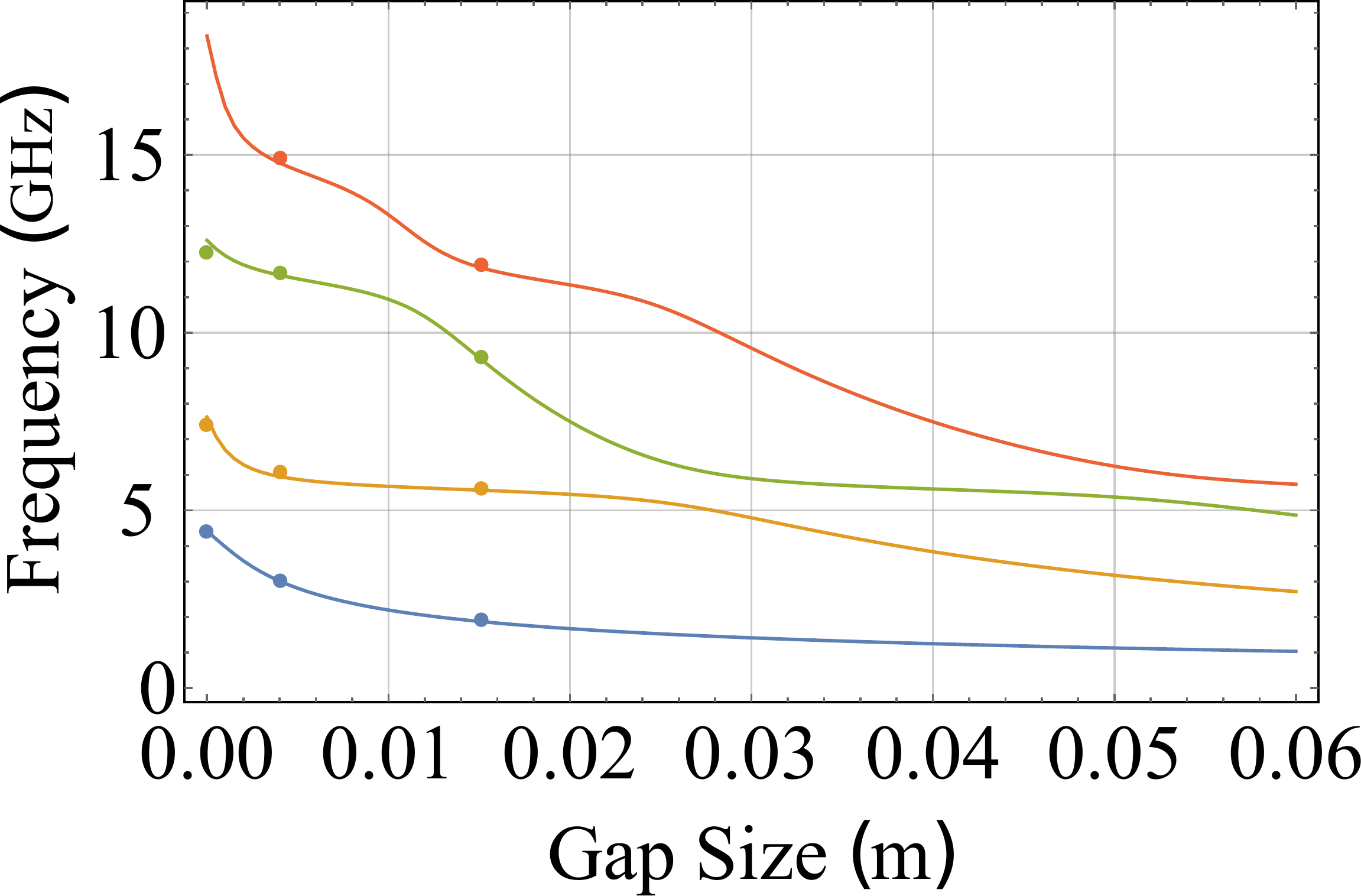}~
	\includegraphics[width=0.9\columnwidth]{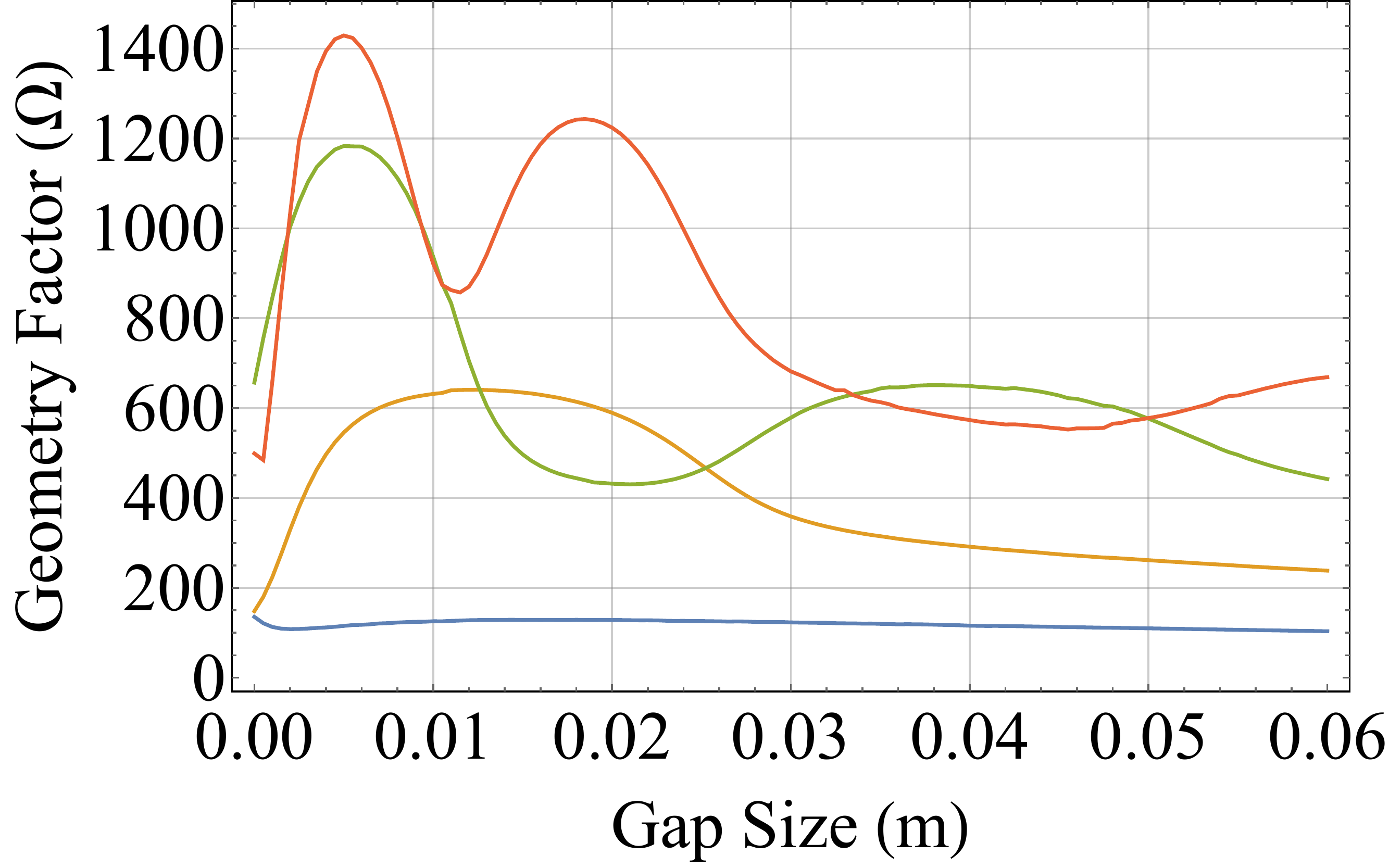}
	\caption{Resonant frequencies (left) and geometry factors (right) as a function of the size of the gap between the sapphire ring and the cavity wall, computed via finite element analysis for the first four TM modes. TM$_{010}$ is shown in blue, TM$_{020}$ in orange, TM$_{030}$ in green, and TM$_{040}$ in red. The sapphire ring was of fixed dimensions (outer radius 24.41 mm, thickness 3.24 mm and height 31.98 mm). The TM modes become Bragg confined modes in the regions where the Geometric Factor is enhanced and the field is anti-resonant in the reflectors. These regions can be seen to be quite broad over gap size variations. Experimental measurements of mode frequencies from cryogenic proof of concept measurements are overlaid on the left plot.}
	\label{fig:BraggExp}
\end{figure*}

The previous super-mode scheme based on disks has a reduced form factor and volume due to much of the in-phase field being present in the dielectric (see eqn. \ref{eq:C}). To improve on this and confine most of the energy in a larger volume and in free space we propose another super-mode scheme that relies on hollow dielectric cylinders of carefully selected thickness and located correctly within a cylindrical conducting shell. This setup is similar to a Bragg resonator, but requires different conditions to achieve optimal axion-sensitivity, so we name them ``Dielectric Boosted Axion Sensitivity", or DBAS resonators.

In a traditional Bragg resonator a dielectric layer inside a cavity creates a virtual boundary condition and thus increases the geometry factor of the resonance, as the mode electromagnetic field is shielded from the comparatively lossy conducting cavity walls. This effect is well documented in the literature for a variety of Transverse Electric modes, and is used to boost quality factors as all wall losses may be decoupled from the mode simultaneously~\cite{Floch:2007aa,Floch:2008aa,Tobar:2001aa,Tobar:2004aa,Krupka:2005aa}. For this effect to work the electric field pattern of the mode must be tangential to the cavity walls. Thus, to apply it to an axion sensitive TM mode only the cylindrical walls can make use of the Bragg effect. This may be achieved by increasing the radius of the cavity slightly, and placing a dielectric boundary inside. To our knowledge this has not been achieved before with TM modes. 

When the correct parameters, such as the dielectric thickness and the size of the gap between the dielectric and the conducting walls are met, the geometry factor of the resonance increases dramatically, the frequency decreases slightly, and the axion form factor decreases, due to the increase in overall volume. A representation of the z-direction electric field of a TM$_{020}$ mode (again computed via finite element analysis) confined by a sapphire cylinder such that the Bragg condition is met, is shown in fig.~\ref{fig:ClassicalBragg}. The geometry factor in this case is increased by a factor of $\sim$1.4 when compared with an empty TM$_{020}$ resonator at the same frequency. This effect may be applied to any mode in the TM$_{0n0}$ family, however, the location and dimensions of the ring relative to the outer wall of the cavity for optimal geometry factor will be different for each mode.

\subsubsection{Bragg resonators for TM modes}
We investigated this effect for different TM modes via finite element modelling in COMSOL Multiphysics, and performed proof of concept measurements. For fixed sapphire ring dimensions (outer radius 24.41 mm and thickness 3.4 mm) we varied the size of the gap between the walls of the resonator and the interior sapphire ring, computing the geometry factors and frequencies of the first four TM modes. The results of this finite element modelling, with overlayed experimental results and including measured quality factors are shown in fig.~\ref{fig:BraggExp} and table.~\ref{table:BestQs}. Whilst such resonators boast increased quality factors compared with traditional TM modes, they are not the most axion sensitive resonant structures that can be constructed with dielectric rings.
\begin{table}[h]
\begin{tabular}{|c|c|c|}
	\hline
	Mode&Quality Factor&Frequency\\
	&(sapphire ring)&(sapphire ring) (GHz)\\
	\hline
	TM$_{010}$&25028&1.92\\
	TM$_{020}$&75086&5.63\\
	TM$_{030}$&98600&11.694\\
	TM$_{040}$&63120&14.96\\	
	\hline
\end{tabular}
	\caption{Measured quality factors for the first four TM modes from proof of concept experiments.}
	\label{table:BestQs}
\end{table}

\begin{figure}\centering
	\includegraphics[width=0.85\columnwidth]{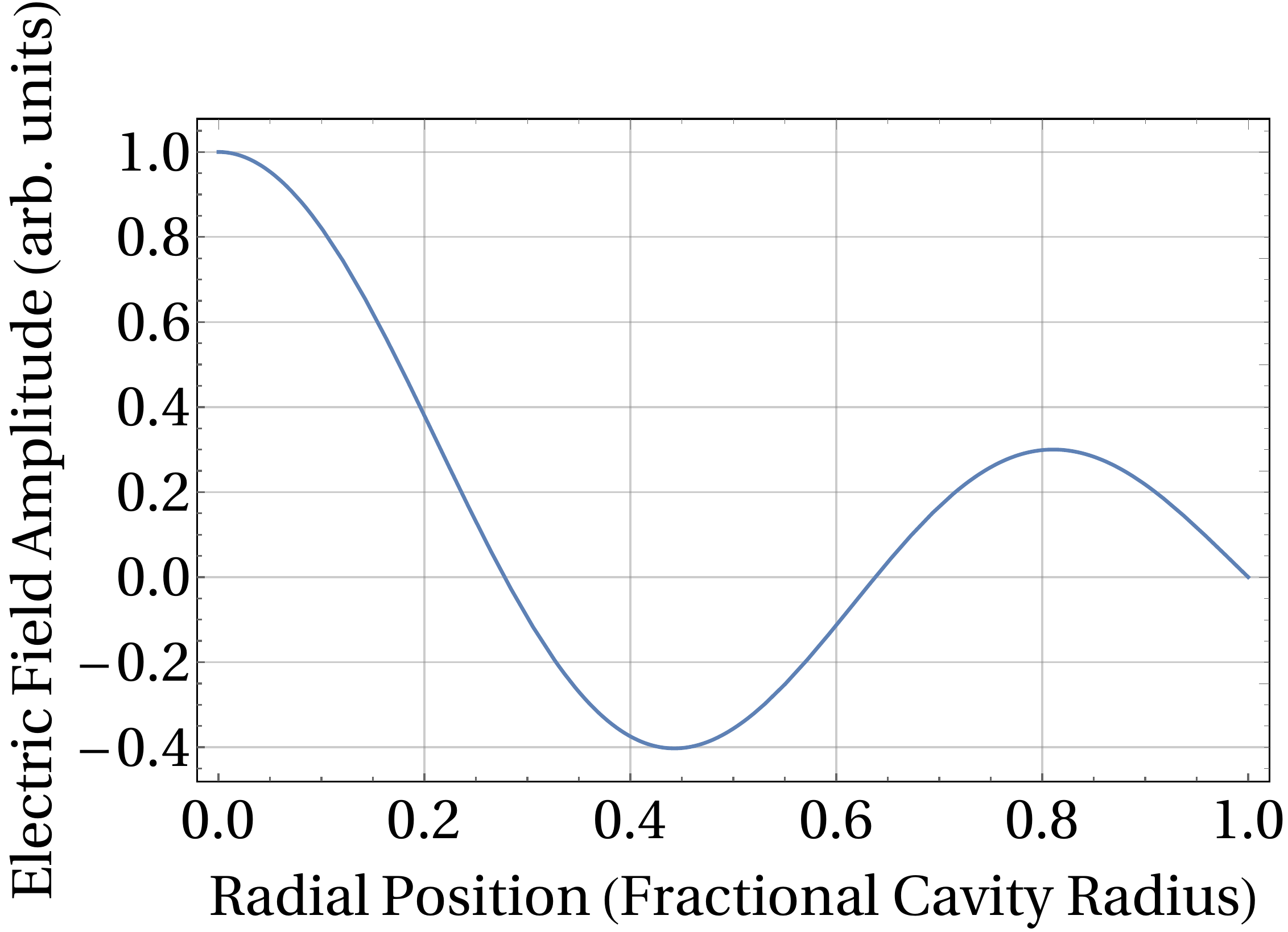}
	\caption{Representation of the $E_z$ component of the TM$_{030}$ mode as a function of radial distance from the centre of the cavity. The mode takes the form of a Bessel J function.}
	\label{fig:BesselJEz}
\end{figure}
\subsubsection{Dielectric Boosted Axion Sensitivity, or DBAS resonators}
\begin{figure*}\centering
	\includegraphics[width=0.8\textwidth]{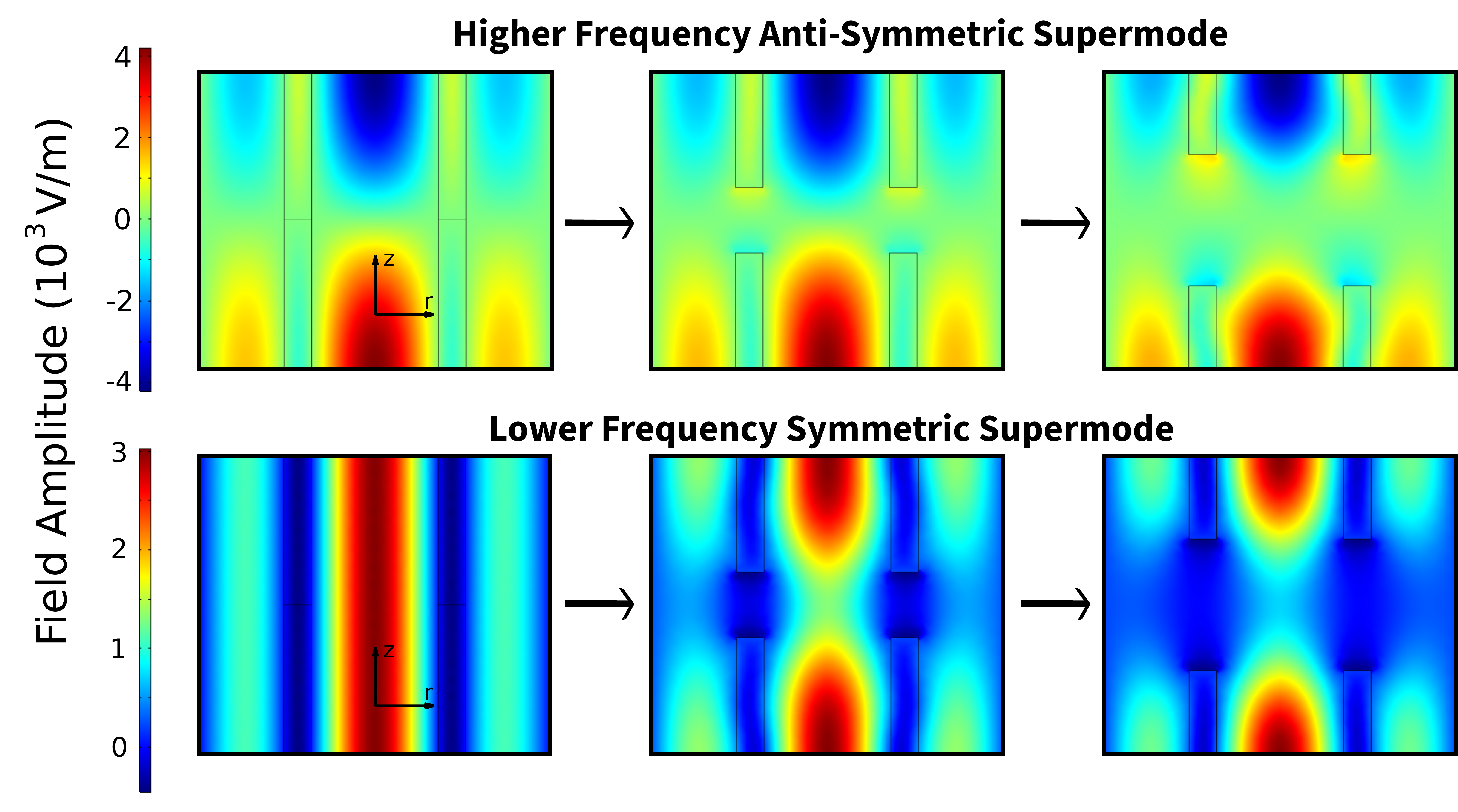}
	\caption{$E_z$ field distribution for the two modes discussed in the text as the gap between the sapphire rings increases. Redder colours represent higher positive $E_z$ values, whereas bluer colours represent higher negative field values. The lower mode is a TM$_{030}$-like mode, with high axion sensitivity, whereas the higher mode is a TM$_{031}$ like mode with no axion sensitivity.}
	\label{fig:TM030Braxion}
\end{figure*}
\begin{figure*}\centering
	\includegraphics[width=0.88\columnwidth]{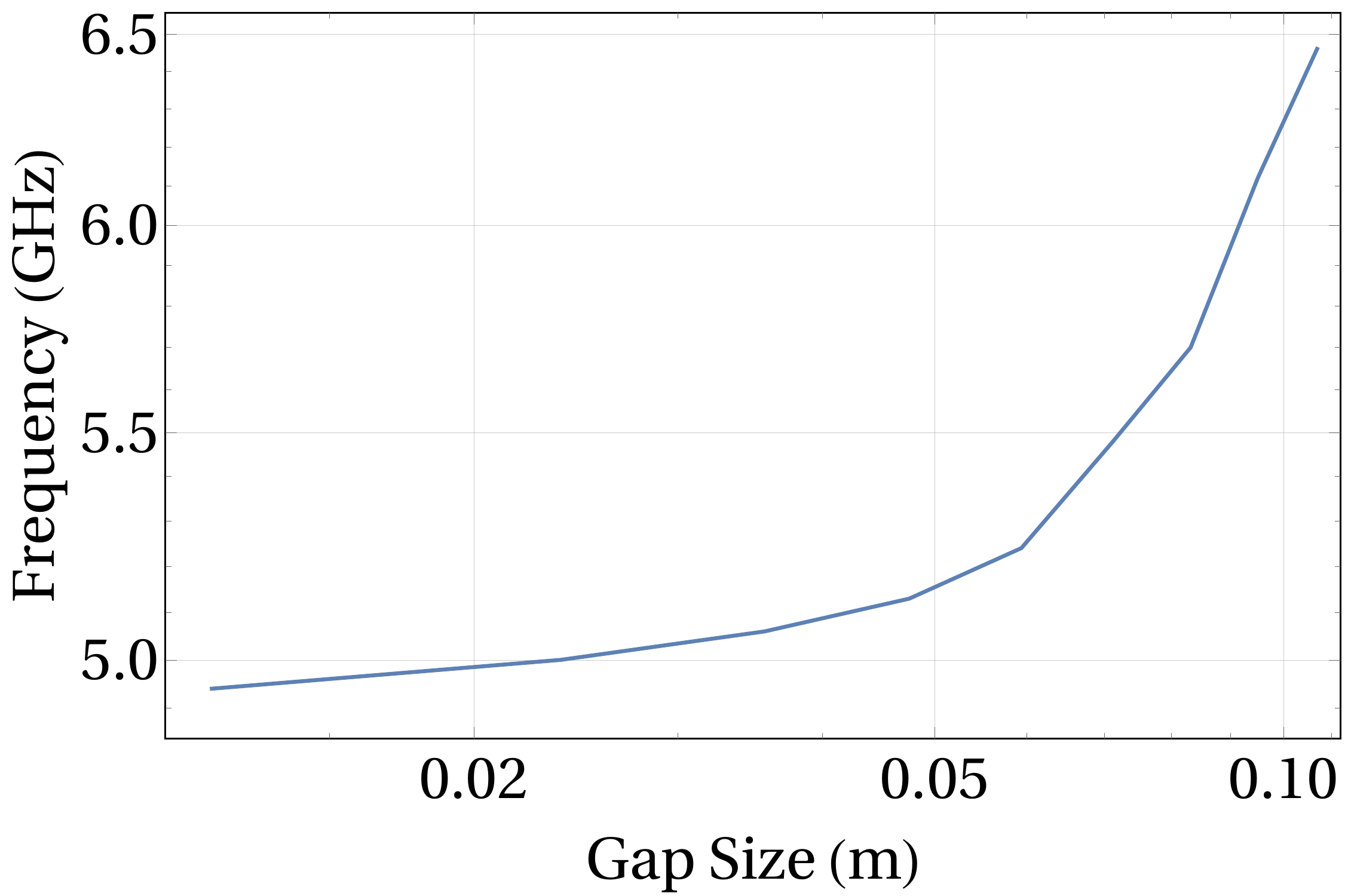}~
	\includegraphics[width=0.88\columnwidth]{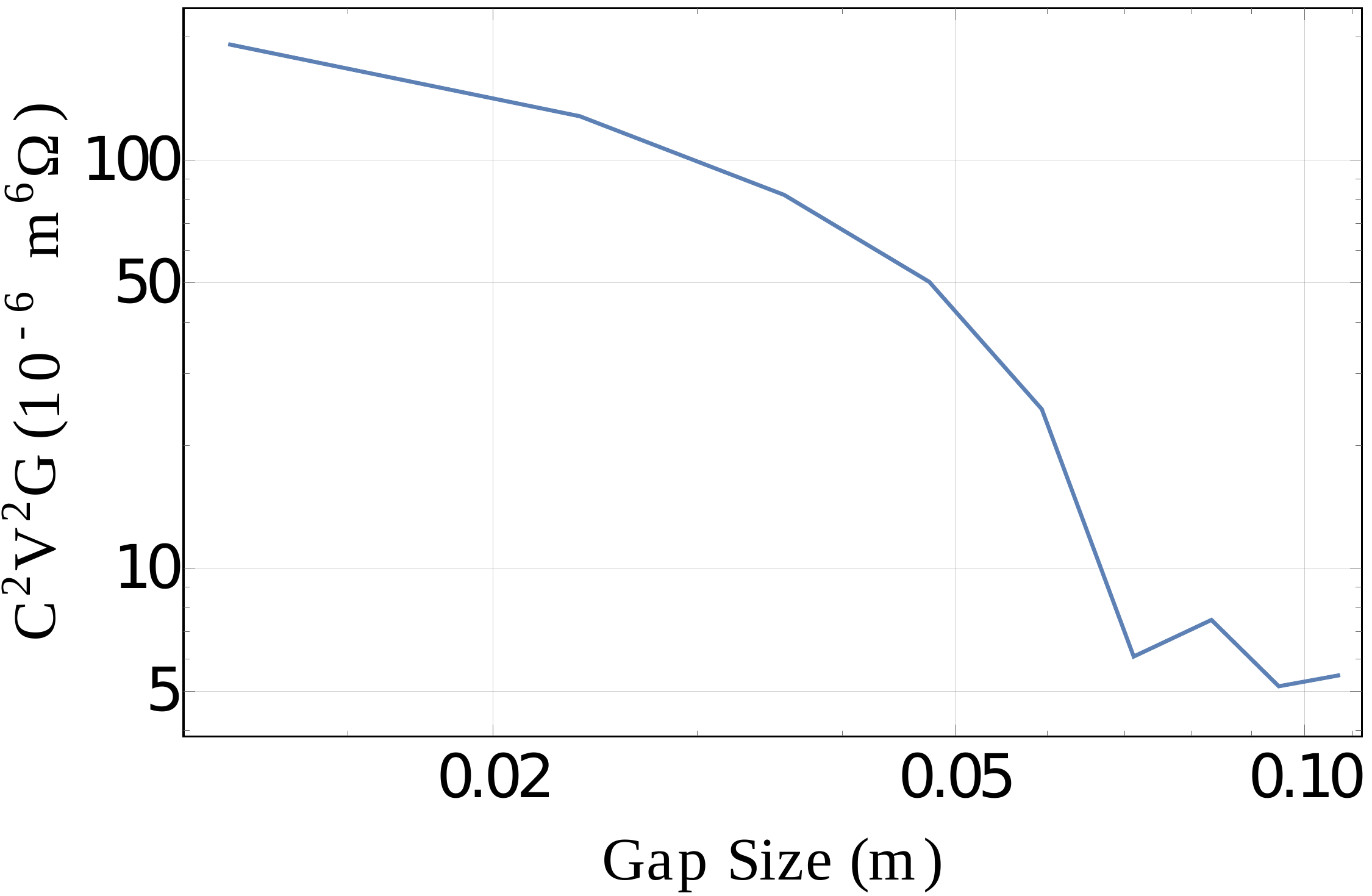}
	\caption{Frequency in GHz (left), and $C^2V^2G$ product (right) vs gap size in metres for the ideal DBAS resonator discussed in the text.}
	\label{fig:BraxionC2V2G}
\end{figure*}
As discussed previously, it is possible to use dielectrics to alter the axion coupling to a cavity resonance, by changing the field structure within the cavity. Consider that for a TM$_{0n0}$ mode inside a cylindrical resonator
\begin{align*}
\vec{E}_c(r)&=E_0~e^{i\omega t}~J_0(\frac{\zeta_{0,n}}{R}r)~\hat{z}\\
R&=\frac{\zeta_{0,n}c}{\omega},
\end{align*}
where $\vec{E}_c$ is the mode electric field, $E_0$ is a constant related to the amplitude of the field, $\omega$ is the mode angular resonant frequency, $J_0$ is the zeroth order Bessel J function, $\zeta_{0,n}$ is nth root of the zeroth order Bessel J function, $R$ is the cavity radius, $r$ is the radial distance from the center of the cavity, $\hat{z}$ is the cavity z-direction unit vector and c is the speed of light. The z-component of electric field takes the form of a Bessel J function and alternates in phase accordingly, as illustrated for a TM$_{030}$ mode in fig.~\ref{fig:BesselJEz}. Upon inspection of equation~\ref{eq:C}, it is clear that the form factor would increase if a higher proportion of the z-component of the electric field was in the same phase, as the numerator of the first term in the expression would increase. In order to achieve this, we can place a dielectric ring with the correct thickness and in the correct position such that the out of phase electric field is confined in the dielectric, and thus its contribution to the form factor is reduced. This technique will not work for a TM$_{010}$ mode, where the field is all in the same direction, and tends to work best for odd numbered modes. To illustrate this we shall consider a TM$_{030}$ mode. We know that the radial distance from the centre of the resonator to the point where the field changes direction is given by
\begin{equation}
r=\frac{\zeta_{0,1}}{\zeta_{0,3}}R\sim0.278~R
\label{eq:InPhase1}
\end{equation}
We should leave this region as vacuum, as we do not wish to reduce the contribution of this section of the field to the form factor. We also know that the radial distance from the centre to the point where the field changes direction for the second time, ie comes back into phase with the first region is given by
\begin{equation}
r=\frac{\zeta_{0,2}}{\zeta_{0,3}}R
\nonumber\end{equation}
So, we conclude that the region containing electric field out of phase with the other two regions is of width
\begin{equation}
\frac{\zeta_{0,2}-\zeta_{0,1}}{\zeta_{0,3}}R\sim0.36~R,
\label{eq:OutOfPhase}
\end{equation}
beginning at $r=\frac{\zeta_{0,1}}{\zeta_{0,3}}R$. The final vacuum region of in phase field will be of thickness
\begin{equation}
\frac{\zeta_{0,3}-\zeta_{0,2}}{\zeta_{0,3}}R\sim0.362~R.
\label{eq:InPhase2}
\end{equation}
The middle, out of phase region in eq.~\ref{eq:OutOfPhase} is the region in which we would like to place a dielectric. Considering the speed of light is reduced in a dielectric medium by a factor of $\sqrt{\epsilon_r}$ we can consider the space inside the dielectric as being effectively increased by this same factor. Thus, to meet our condition that the out of phase field be confined in the medium, we construct our dielectric region such that its size is reduced compared with eq.~\ref{eq:OutOfPhase} by the same factor. Put simply, our dielectric ring should have thickness
\begin{equation}
\frac{\zeta_{0,2}-\zeta_{0,1}}{\zeta_{0,3}\sqrt{\epsilon_r}}R\sim0.107~R,
\label{eq:OutOfPhaseReduced}
\end{equation}
for sapphire, $\epsilon_r\sim11.349$. Now of course, as the middle region is reduced in size by a factor of $\sqrt{\epsilon_r}$, which is over 3 for sapphire, these fractions of $R$ (0.278, 0.107, and 0.362, eqs.~\ref{eq:InPhase1},\ref{eq:InPhase2},\ref{eq:OutOfPhaseReduced}) do not sum to unity, but are rather fractions $\emph{relative}$ to one another. That is to say, the ratio of the first vacuum region thickness to the sapphire region thickness should be $\frac{0.278}{0.107}$. In order to simplify this, we can multiply all of the relevant thicknesses by $\frac{1}{0.278+0.107+0.362}\sim1.34$ such that they sum to unity. Specifically, the regions should be of thickness $0.372~R,~0.143~R,$ and $0.485~R$ respectively.

As an aside, this increase in size makes intuitive sense, as the introduction of the dielectric reduces the frequency of the cavity slightly, and the ratio by which we multiply the values is almost exactly equal to the ratio of the radius of an equivalent frequency empty cavity, to the radius of the cavity containing the dielectric ring. We may equivalently think of it as follows: the regions should be of the thicknesses presented in eqs.~\ref{eq:InPhase1},\ref{eq:InPhase2},\ref{eq:OutOfPhaseReduced}, if we replace $R$ in these equations by $R'$, the radius of a larger cavity increased in size relative to the cavity containing the dielectric ring by a factor of $\sim1.34$ (for sapphire), such that the frequency of this larger empty cavity is the same as that of the smaller cavity containing the ring. Consider $R'$ the \emph{effective} radius of the cavity, due to the effect of the dielectric.

Whilst hopefully illustrative, this is not very helpful in designing a cavity, as we do not know what the final frequency of the ring cavity is going to be until we have decided on a ring thickness, which by this method we cannot find until we know the radius $R'$ of the larger equivalent cavity, which we in turn cannot find until we know the final ring cavity frequency. We should instead opt for the ``sum to unity" method outlined above, and employ thicknesses of $0.372~R,~0.143~R,$ and $0.485~R$ for the first vacuum, dielectric, and second vacuum region respectively (for sapphire, in a TM$_{030}$ mode, this same process may be followed for other materials in other modes). These are the dimensions for the air and sapphire layers that should be chosen if we wish only to constrain the out of phase field within the dielectric, and maximize our axion cavity form factor. This method can be easily extended following the above methodology to any odd n-valued TM$_{0n0}$ mode, with $\frac{n-1}{2}$ layers of dielectric. The TM$_{030}$ mode is presented here as it is the most straightforward, and most experimentally practical to implement, requiring only one dielectric ring.

We have conducted finite element modelling of a TM$_{030}$-like resonator with a sapphire ring placed to satisfy the conditions above, and can report form factors on the order of 0.47, compared with 0.053 for an empty cavity TM$_{030}$ mode. Furthermore, the geometry factor with the correctly placed sapphire was 791, compared with 269 for an empty TM$_{010}$ cavity of the same frequency and aspect ratio. A representation of the z-direction electric field in this case is shown in fig.~\ref{fig:TM030Braxion}.

Interestingly, we can exploit a super-mode type tuning effect with this structure in a similar fashion to the disk structure discussed above. If we split the dielectric ring in the axial direction we observe that there are again two modes, the symmetric TM$_{030}$ mode occupying the entire dielectric ring, and a higher frequency anti-symmetric mode which contains a single variation in the longitudinal direction, a TM$_{031}$ mode. If we increase the gap size between the two parts of the ring we observe that the frequency of the axion sensitive symmetric super-mode, TM$_{030}$, increases whilst retaining good sensitivity. For a cavity with a radius of 63.8 mm, a height of 108 mm, a sapphire ring of thickness 9.12 mm, with a TM$_{030}$-like starting frequency 4.83 GHz, we observe tuning of approximately 1.6 GHz as the ring is separated and removed from the cavity. The $C^2V^2G$ product and resonant frequencies for this resonator, computed via finite element analysis is shown in fig.~\ref{fig:BraxionC2V2G}. This super-mode tuning scheme works the same way for a traditional Bragg resonator, or the DBAS resonator. To our knowledge this is the first time that a super-mode tuning method has been applied to Bragg or Bragg-like resonators, and we stress that this or similar techniques could be applied to other Bragg resonant structures in other applications, to create highly tunable resonators. For example, the same ideas could be applied to high-Q TE modes based on Bragg reflectors, to create highly tunable cavities, whereas traditional Bragg resonators are stationary in frequency.

These dimensions were chosen so that the resonator could be compared to the sapphire disk resonator. It is important to note that, similar to tuning the sapphire disk resonator, there are practical considerations in the design of a DBAS effect resonator. In order to reduce complexity we may again choose to displace only one of the two rings inside the cavity, adjusting the gap between them. In such a case it is optimal to place the break in the ring below the half way point,  as this provides more symmetry over the range of tuning. With the break in the middle, the only point of axial symmetry is at 0 gap, whereas with the cut below the middle there are two points of axial symmetry over the tuning range.

It is worth noting that, whilst there are mode crossings with azimuthally varying (m~$>0$) modes, finite element analysis did not predict avoided level crossings and associated reduction in sensitivity. We believe that this is a result of a orthogonality between these higher order modes and the axion sensitive m=0 mode. Consequently it would be possible to employ the standard technique of filling the resonator with a different material, such as liquid helium, and rescanning over the regions where the sensitive m=0 mode crosses these higher order modes~\cite{Daw:1998jm}.

Resonators of this type are very useful in the push towards higher frequency haloscopes where it is beneficial to utilize higher order modes, so that larger cavities may be employed. Traditionally this is not appealing as the form factors for these higher order modes are significantly reduced such that the product $CV$ remains nearly constant. We have presented a scheme for utilizing large resonators with higher order modes, whilst still maintaining appreciable form factors and high axion sensitivity. The scheme also comes with a ``built-in" tuning mechanism which is highly responsive to position displacement based on super-mode tuning.

\begin{figure}
	\includegraphics[width=0.9\columnwidth]{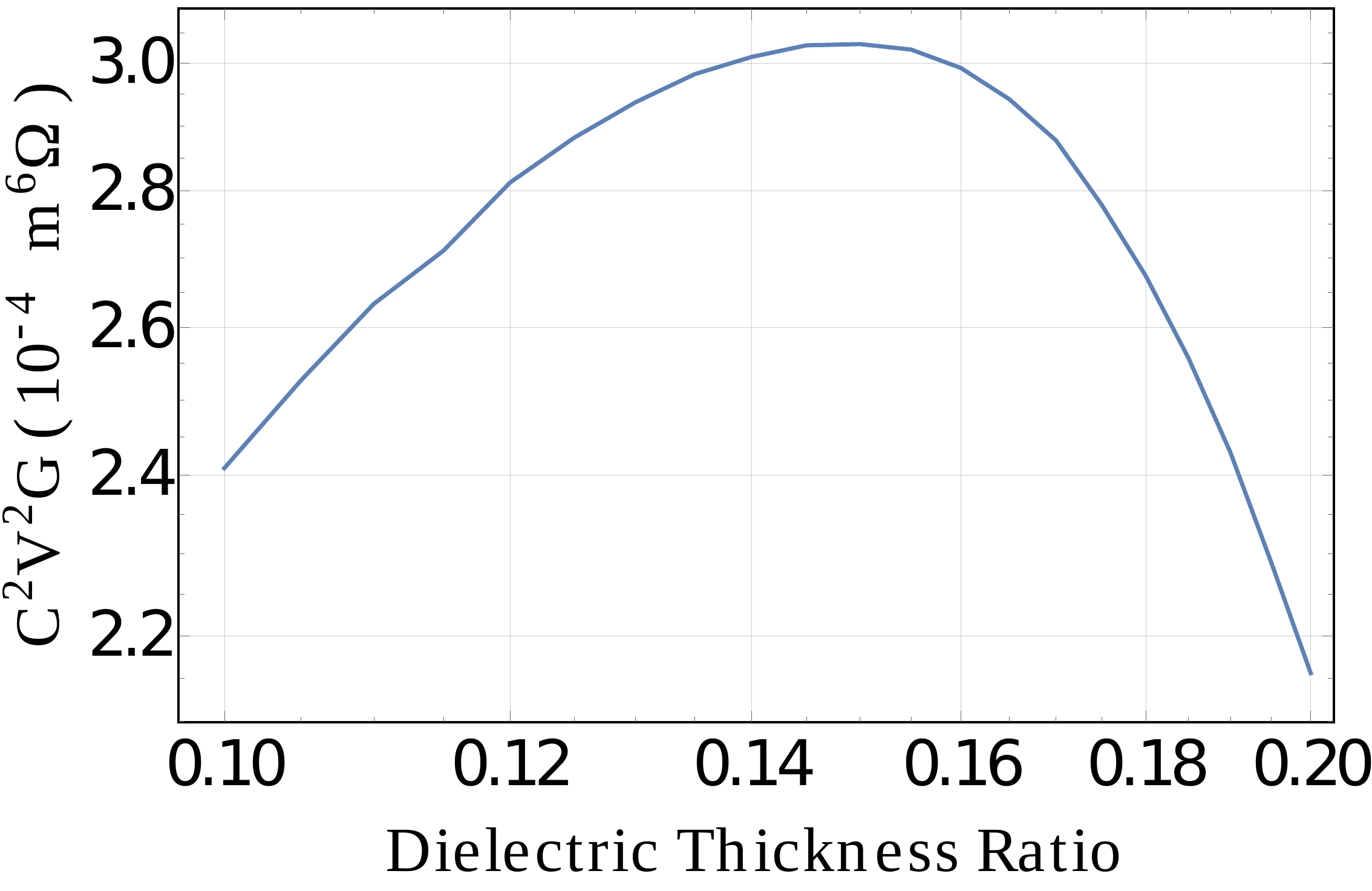}
	\caption{$C^2V^2G$ product versus thickness of the dielectric region as a fraction of cavity radius, with each point scaled to 5 GHz for comparison. The gap size and other parameters were kept constant in this sweep. As shown the optimum lies near 0.15, which is close to what is predicted in the text for the DBAS effect.}
	\label{fig:Optimize}
\end{figure}

It is important to note that the optimal condition for geometry factor (Bragg effect) and the optimal condition for form factor (DBAS effect) are not the same, and consequently a trade-off may be required. The optimal $C^2V^2G$ product may occur for slightly different dielectric thickness and location depending on the specific parameters of the cavity and experiment, such as length, mode frequency, and desired tuning range. In order for a shift away from the DBAS condition to be optimal, we must be gaining in geometry factor much faster than we are losing in form factor (as the figure of merit depends on form factor quadratically, but only linearly in geometry factor), but this may be possible depending on the specific geometry of the resonator under design, and the relative narrowness or broadness of the Bragg and DBAS regions. This is very complex to model analytically, and so in order to design resonators for a haloscope based on these techniques, we employ an iterative method where we adjust dielectric thickness and location (starting at the DBAS result, and making small changes) and compare results until the optimal parameters are found. An example of the results of this kind of iterative process are shown in fig.~\ref{fig:Optimize}. In any case, we may wish to call the optimal resonator (which may be a hybrid Bragg and DBAS resonator) a ``Bragg-Axion" or ``Braxion" reosnator.
\begin{figure*}\centering
	\includegraphics[width=0.85\columnwidth]{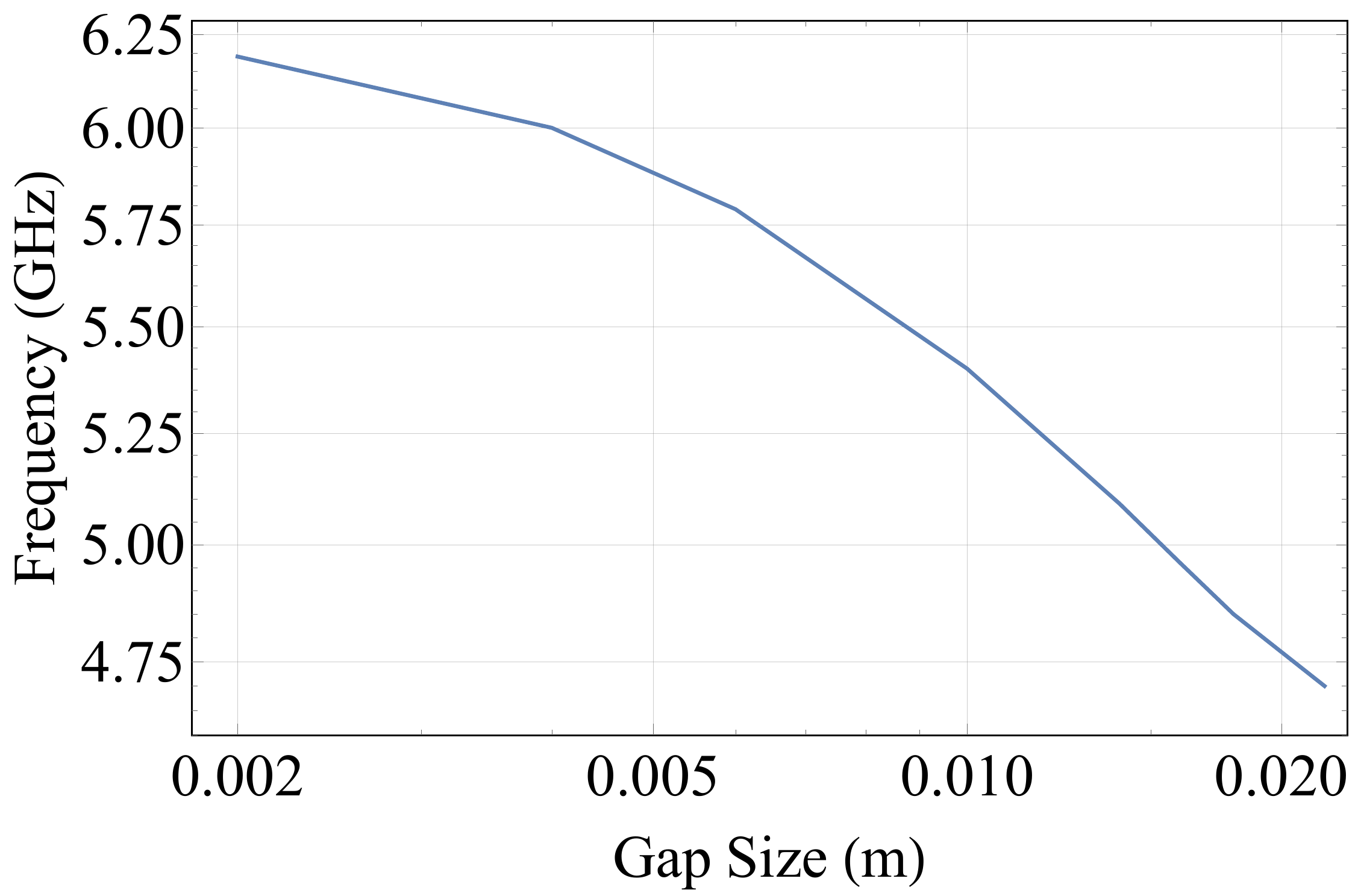}~
	\includegraphics[width=0.85\columnwidth]{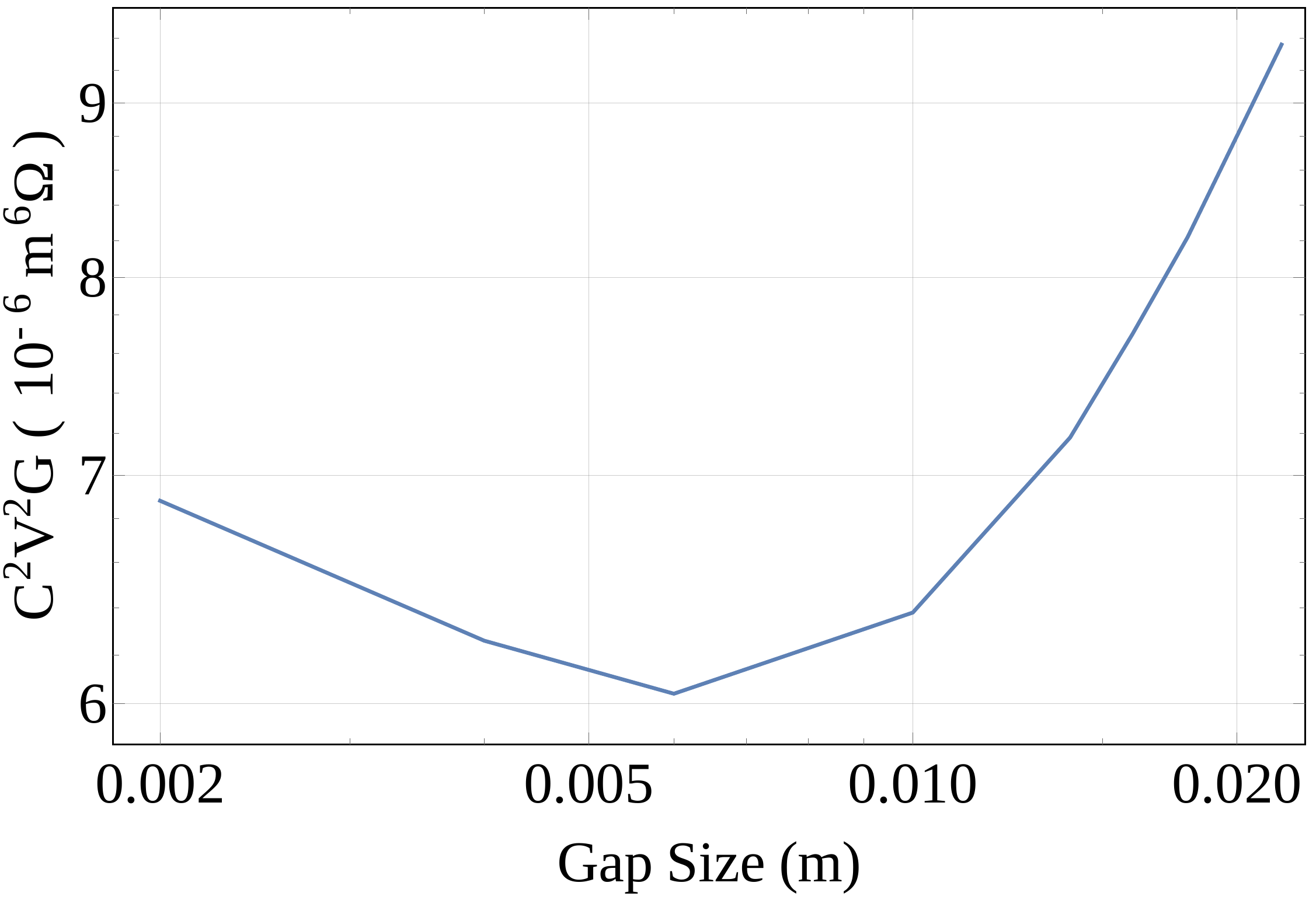}
	\caption{Frequency in GHz (left), and $C^2V^2G$ product (right) vs rod distance from the centre of the cavity in metres for the conducting rod-tuned resonator discussed in the text.}
	\label{fig:c2v2gtypical}
\end{figure*}
\subsection{Comparison with traditional tuning rod}
It is a common technique for tuning a haloscope to introduce a conducting rod into the cavity, and slide it radially from the outer region of the cavity into the centre. This breaks symmetry in the TM$_{0n0}$ mode family and alters the frequency of the cavity. We present a $C^2V^2G$ plot, and tuning range for such a cavity (fig.~\ref{fig:c2v2gtypical}), employing a TM$_{010}$ mode, computed via finite element analysis. This cavity was designed such that the tuning range  and frequencies were comparable to the two schemes presented in this article. It is clear from the results that the DBAS resonator scheme proposed here outperforms the traditional tuning rod in a TM$_{010}$ cavity around 5 GHz. The DBAS resonator has $C^2V^2G$ products that are, at their peak, between 1 and 2-orders-of-magnitude greater than the traditional tuning rod. The dielectric disk scheme has $C^2V^2G$ products that are lower by an order of magnitude, however, the resonator tunes extraordinarily quickly and spurious mode density in the neighbourhood is low. This type of resonator would be of interest in an experiment that required a very fast scanning rate, such as those proposed to search for streaming dark matter~\cite{StreamingDM}. We consider these, particularly the DBAS resonator, to be valuable resonant structures for axion haloscopes, and such resonators will be employed in the forthcoming ORGAN experiment~\cite{ORGANPDU}.

\section{Conclusion}
Two novel dielectric resonator designs for axion haloscopes have been presented. Additionally the concept of a Bragg resonator was applied to TM modes showing increased quality factors, rather than the usual TE modes for such structures. Both haloscope  schemes have ``built-in" tuning mechanisms highly responsive to position displacement based on super-mode interactions. The first haloscope scheme confines most of the field in the dielectric, whilst the second confines only the lesser, out of phase field components in the dielectric. This work also represents the first application of both Bragg resonators and super-mode tuning to axion haloscopes. The advantages of these schemes in regards to different axion searches were discussed. The microwave cavity Bragg and Dielectric Boosted Axion Sensitivity conditions are discussed and compared, as they are different. Results of finite element analysis have allowed the optimisation of the sensitivity and scan rate based on the figure of merit, $C^2V^2G$. We have undertaken a proof of concept experiment for the dielectric disk resonator scheme, as well as for the Bragg effect in TM modes, showing good agreement with the modelling. Finally, we compare these two resonator schemes with a traditional conducting rod-tuned haloscope and find that the second scheme in particular, the DBAS resonator is far superior.

This work was funded by Australian Research Council (ARC) grant No. CE170100009, the Australian Government's Research Training Program, and the Bruce and Betty Green Foundation.
%

\end{document}